\documentclass[aps,prd,showpacs,nofootinbib,superscriptaddress,amsmath,amssymb]{revtex4}
\usepackage{graphicx}
\usepackage{amsmath}
\usepackage{amssymb}

\usepackage{amsfonts}
\usepackage{bm}
\usepackage{verbatim}

\usepackage{graphicx}
\usepackage{graphics}

\usepackage{color}

\usepackage{epsfig}

\begin{document}

\title{The black hole interior and the type II Weyl fermions}

\author{M.A. Zubkov \footnote{On leave of absence from Institute for Theoretical and Experimental Physics, B. Cheremushkinskaya 25, Moscow, 117259, Russia}}
\email{mikhailzu@ariel.ac.il}
\affiliation{Physics Department, Ariel University, Ariel 40700, Israel}


\begin{abstract}
It was proposed recently that the black hole may undergo a transition to the  state, where inside the horizon the Fermi surface is formed that reveals an analogy with the recently discovered type II Weyl semimetals. In this scenario the low energy effective theory outside of the horizon is the Standard Model, which describes excitations that reside near a certain point $P^{(0)}$ in momentum space of the hypothetical unified theory. Inside the horizon the low energy physics is due to the excitations that reside at the points in momentum space close to the Fermi surface. We argue that those points may be essentially distant from $P^{(0)}$ and, therefore, inside the black hole the quantum states are involved in the low energy dynamics that are not described by the Standard Model. We analyse the consequences of this observation for the physics of the black holes and present the model based on the direct analogy with the type II Weyl semimetals, which illustrates this pattern.
\end{abstract}

\maketitle

\section{Introduction}

Soon after the discovery of the Schwarzschild black hole solution \cite{Schwarzschild}, its form was proposed, which allows to reveal an analogy with the motion of liquid \cite{Gullstrand,Painleve}. The corresponding coordinate system is called Painleve - Gullstrand coordinate frame. Later the similar reference frame was proposed for the charged (Reissner-Nordstrom) and even for the rotated (Kerr) black holes \cite{Hamilton:2004au,Doran:1999gb}. The simulation of the black holes by real motion of the superfluid was proposed in \cite{Volovik:1999fc}. In the same
 paper the semiclassical calculation of Hawking radiation \cite{Hawking:1974sw} was proposed for the first time. The similar calculation was later reproduced in \cite{Parikh:1999mf}, which was followed by a number of papers (see, for example, \cite{Akhmedov:2006pg,Jannes:2011qp} and references therein).
The clear formulation of the analogy between the black holes (including rotated and charged ones) and the fluid motion is given in \cite{Hamilton:2004au}. Within this pattern in \cite{VolovikBH} it has been proposed that the conventional state of the black holes (that is accompanied by Hawking radiation) may be transformed to the different, equilibrium state, in which the interior of the black hole contains the Fermi surface.

\begin{figure}
\begin{center}
 \epsfig{figure=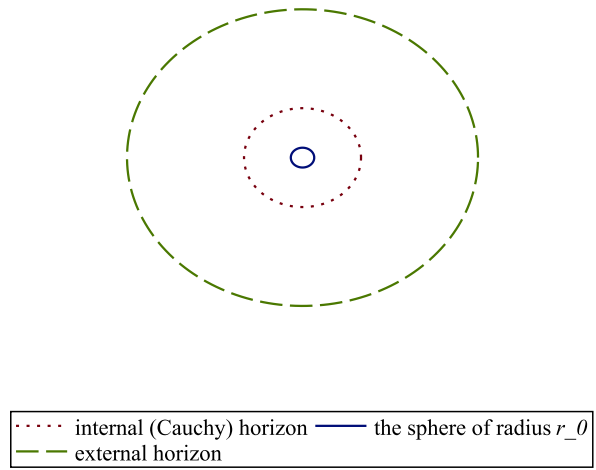,width=80mm,angle=0}
\end{center}
\caption{\label{fig}  Charged non - rotated black hole in Painleve – Gullstrand reference frame is represented schematically.}
\end{figure}

Quite unexpectedly the similar pattern appears in the condensed matter physics. First of all, as it was mentioned above, there exists the analogue of the black hole in fermionic superfluids (see also \cite{Volovik2003}). Moreover, the simulation of motion in the presence of the gravitational field and even the  black hole analogues were discussed in the framework of ordinary liquids \cite{liquid}. Recently the new classes of solids called Weyl and Dirac semimetals were discovered \cite{semimetal_effects6,semimetal_effects10,semimetal_effects11,semimetal_effects12,semimetal_effects13,Zyuzin:2012tv,tewary,16}. In those materials the low energy excitations are described by the same Dirac (or, Weyl) equation as the elementary particles of high energy physics. This allows to simulate in laboratory the effects typical for the high energy physics. Especially interesting is the possibility to observe effects, which were not observed in experimental high energy physics due to the restrictions on the energy of the corresponding processes.

\begin{figure}[!thb]
\begin{center}
{\epsfig{figure=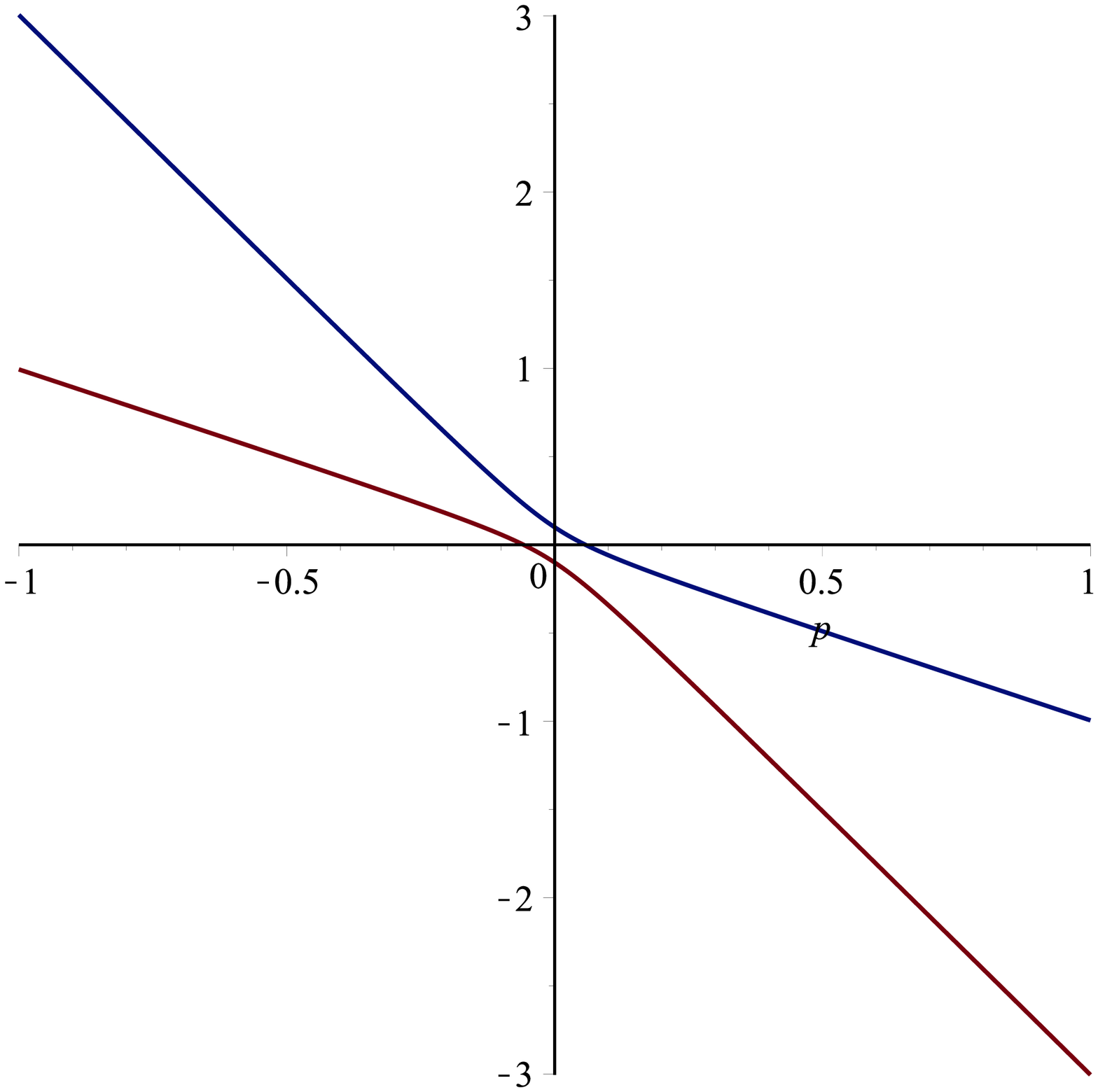,width=40mm,angle=0}}
{\epsfig{figure=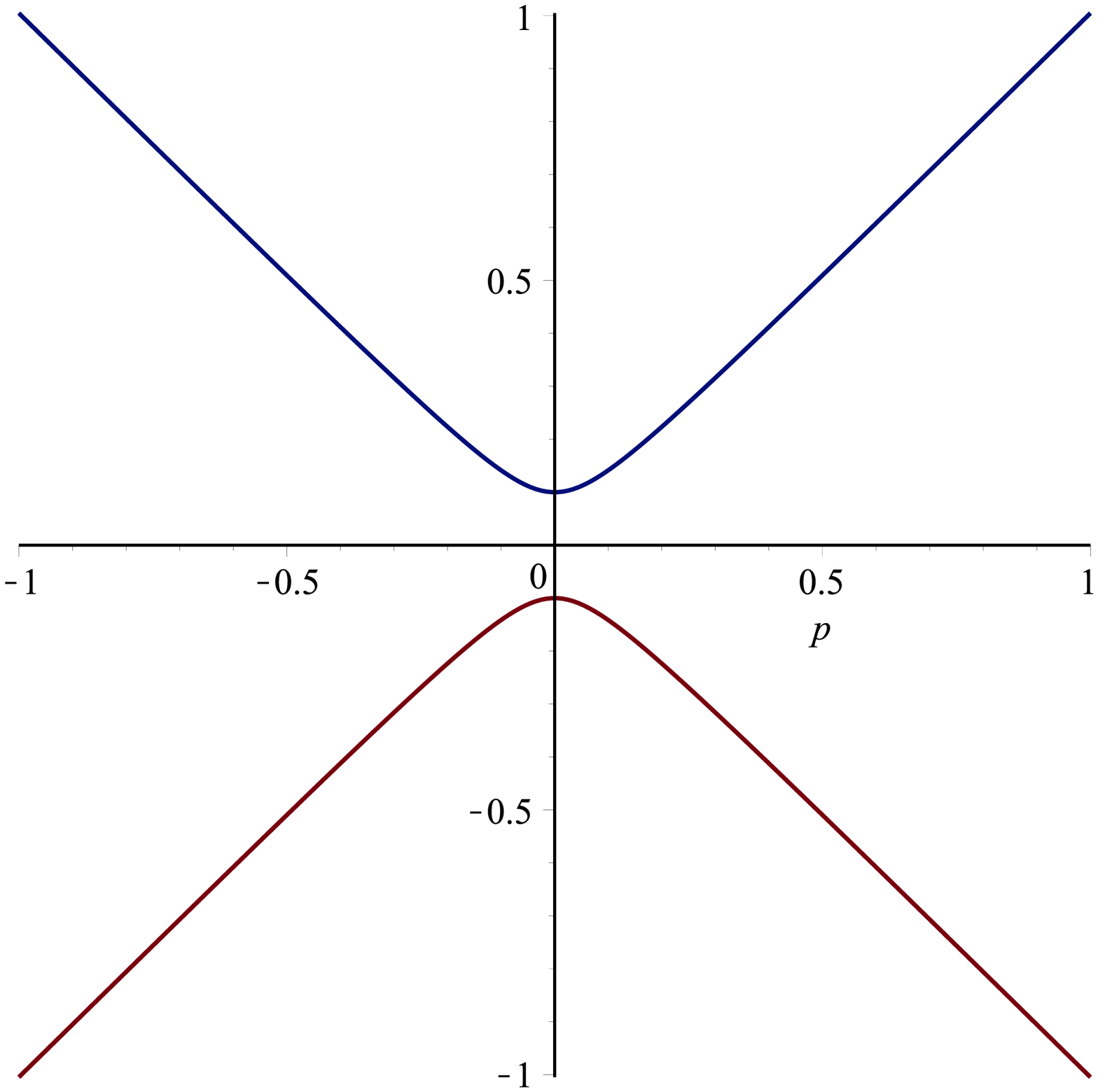,width=40mm,angle=0}}
\end{center}
\caption{\label{fig.0}  {\bf Left:} Dispersion for the Dirac fermions inside the black hole horizon (in the Painleve - Gullstrand reference frame). The lower line symbolizes the occupied branch of spectrum. One can see, that its part is situated above zero, which results in the tunneling of the occupied "vacuum" states from the interior of the black hole to its exterior.  {\bf Right:} Dispersion for the Dirac fermions outside of the black hole horizon. The lower line symbolizes the occupied branch of spectrum. It is situated below zero.}
\end{figure}

In \cite{VZ} (Remark 2.1.) the possible existence of the new type of Weyl points in multi - fermion systems of general type was noticed. Later the corresponding materials were called the type II Weyl semimetals and discovered experimentally \cite{W2}. It appears, that in those novel materials the fermionic quasiparticles behave similar to the fermions inside the horizon of the equilibrium black holes discussed in \cite{VolovikBH}. This analogy including the Hawking radiation has been discussed in \cite{VolovikBHW2}. Further, basing on this analogy the more exotic forms of the Weyl fermions were proposed in \cite{NissinenVolovik2017a}. This line of research represents, to a certain extent, the alternative to the more popular approach to the quantum theory of the black holes that is based on the black hole thermodynamics (see, for example, \cite{Padmanabhan:2003gd} and references therein) and on the string theory \cite{Susskind_book}.

By the quantum theory of the black holes we understand here the attempt to apply quantum theory to the physics of the black holes, which therefore are not already considered as completely classical objects contrary to the conventional general relativity.  The review of the present state of the study in the quantum theory of black holes is presented, for example, in \cite{Suuskind}, where more references may be found (see also  \cite{Hamilton:2004aw} and references therein). For the review of the application of quantum theory
to the physics of black holes (including the approaches based on string theory) see \cite{Susskind_book,Mann,information,BHI} and references therein.

The mentioned above alternative line of research in the field of the quantum theory of black holes is based on an analogy with real condensed matter systems (the fermionic superfluids and Weyl semimetals). It was initiated in \cite{Volovik:1999fc} and developed further in \cite{VolovikBH,VolovikBHW2,NissinenVolovik2017a}. The most important advantage of this approach is the possibility to observe in laboratory the phenomena that are specific for the real black holes that may, possibly, exist in the Universe and manifest themselves in astrophysical observations \footnote{It is worth mentioning that recently certain astrophysical observations were interpreted as the indication that the black holes do exist in the real World (see, for example, \cite{experiment_BH}).}. In the mentioned papers the following scenario for the evolution of the black holes was proposed. At the beginning of the formation of the black hole the vacuum is free falling to its center.  This free falling is accompanied by Hawking radiation, which may be detected in any stationary reference frame. And this is the conventional state of the black holes. This free - falling "vacuum" is not stable. Therefore, at the later stages of the existence of the black hole the vacuum may be reconstructed. If equilibrium is achieved, the Hawking radiation stops there, or, better to say, there is the equilibrium between absorbtion and emission of thermal matter by the black hole. It was demonstrated  that in the Painleve - Gullstrand reference frame the equilibrium vacuum inside the horizon is similar to the so - called type II Weyl semimetals \cite{VolovikBHW2}, discovered recently. There the Fermi surface is formed, which is composed of the two Fermi pockets.

In the present paper we proceed this line of research using an analogy with the solid state physics. We
argue, that the self - consistent field theory (which is able to describe the interior of the black hole in its equilibrium) should be different drastically from the conventional Standard Model. Several unusual consequences follow from the simple assumptions about the mentioned theory. We illustrate those features of the theory by the toy lattice model that reveals an analogy to the type II Weyl semimetals.

The paper is organized as follows. In Sect. \ref{Section_DF} we recall the basic notions of Weyl and Dirac fermions in the background of the charged non - rotated black hole considered in Painleve - Gullstrand reference frame. In Sect. \ref{Sect_RE} we discuss how the conventional vacuum is seen in this reference frame and describe in this framework the Hawking radiation. For the completeness we also remind the reader the semiclassical description of the similar process - the Schwinger pair production by the electric field of the black hole. In Sect. \ref{Sect_RE} we reproduce the reasoning of \cite{VolovikBH} and explain why the conventional black hole may undergo the transition to the state, where inside the black hole the Fermi surface appears. In addition we explain in this section why in order to describe the physics inside the black hole the ultraviolet completion of the Standard Model should be considered. In Sect. \ref{Sect_EQ} we illustrate the principal features of the quantum field theory in the presence of the black hole by an example of the toy model defined in lattice regularization.  It is demonstrated, how  in this model the Fermi surface is formed inside the horizon. Next, we discuss the excitations  that exist inside the black hole horizon near to the Fermi surface.  In Sect. \ref{Sect_CON} we end with the conclusions.

\section{Dirac fermions in the background of the Reissner-Nordstrom black hole (Painleve – Gullstrand reference frame)}
\label{Section_DF}


In the Painleve – Gullstrand reference frame the metric of the non - rotating black hole has the form
\begin{equation}
ds^2 = dt^2 - (d{\bf r} - {\bf v}({\bf r}) dt)^2, \label{PGQ}
\end{equation}
where
\begin{equation}
{\bf v} = -\frac{1}{m_P} \frac{\bf r}{r}\, \sqrt{\frac{2M}{r}-\frac{Q^2}{r^2}}
\end{equation}
is the velocity of the free falling body (measured in its self time).
Here $Q$ is the total charge of the black hole, $m_p$ is the Plank mass, $M$ is the mass of the black hole.

The given form of metric corresponds to the following vierbein $E^\mu_a$
\begin{equation}
E^\mu_a = \left(\begin{array}{cc}1 & {\bf v}\\0 & 1\end{array} \right)
\end{equation}
and the inverse vierbein $e_\mu^a$
\begin{equation}
e_\mu^a = \left(\begin{array}{cc}1 & -{\bf v}\\0 & 1\end{array} \right)
\end{equation}
The latter is related to metric $g_{\mu \nu}$ as
$$ g_{\mu \nu} = e^a_\mu e^b_\nu \eta_{ab}$$
where $\eta_{ab} = {\rm diag}\,(1,-1,-1,-1)$.
 The only nonzero component of the spin connection $\omega_{ab \mu}$ is the one with the
 space - time index $\mu$ corresponding to the radial direction, the indices $a,b$ corresponding to time direction and radial direction:
 \begin{equation}
 \omega_{r0r} = -\omega_{0rr} = -\frac{d}{dr}|{\bf v}({\bf r})|\label{omega}
 \end{equation}

The action for the neutral right - handed (left - handed) fermions is given by
\begin{eqnarray}
S_\pm &=& \int d^4x \, {\rm det}^{-1}({\bf E})\,\bar{\psi}_\pm(x) \Big(i  E^0_0\partial_t \pm i E^k_a \tau^a \partial_k  \Big) \psi_\pm(x) \nonumber\\ &=& \int d^4x \, \bar{\psi}_\pm(x) \Big(i \partial_t - H^\pm(-i \partial) \Big) \psi_\mp(x)
\end{eqnarray}
Here the sign plus corresponds to the right - handed fermions while the sign minus corresponds to the left - handed ones.  The Hamiltonians $H^{\pm}$ are given by:
\begin{equation}
H^{\pm}({\bf p}) = \pm {\bf p} \sigma - {\bf p} {\bf v}
\end{equation}
The extra term is to be introduced in order to consider Dirac fermions with mass $m$:
\begin{eqnarray}
S_m &=& - m \sum_\pm\int d^4x \, {\rm det}^{-1}({\bf E})\,\bar{\psi}_\pm(x) \psi_\mp(x)\nonumber\\ &=& - m \int d^4x \,\sum_\pm \bar{\psi}_\pm(x)  \psi_\mp(x)
\end{eqnarray}
It is interesting that the spin connection with the only component of Eq. (\ref{omega}) disappears from the above action.

Metric of the form of Eq. (\ref{PGQ}) possesses the horizon at
\begin{equation}
r_+=\frac{M + \sqrt{M^2 - Q^2 m_P^2}}{m_P^2}
\end{equation}
and the so - called Cauchy horizon at
\begin{equation}
r_-=\frac{M - \sqrt{M^2 - Q^2 m_P^2}}{m_P^2}
\end{equation}
For $ r > r_+$ the quantum field theory contains the usual Weyl fermions, and the type I Weyl points appear at $m=0$. At $r_- < r < r_+$ in the absence of Dirac mass we deal with the two type II Weyl points coinciding at the common position in momentum space. This region inside the black hole horizon corresponds to the values of $|{\bf v}|$ that exceed the speed of light. The maximal value of $|{\bf v}|$ is achieved at $r_m = \frac{Q^2}{M}$.

There is also another special point of the Painleve - Gullstrand space - time
$$r_0 = \frac{Q^2}{2M}$$
For $r < r_0$ metric of the form of Eq. (\ref{PGQ}) contains imaginary velocity $v$. In Sect. II.B in \cite{Hamilton:2004au} this point has been caller the turnaround point, because in the analytical continuation of space - time in the Painleve - Gullstrand reference frame at $ r >  r_0$ the flow of space is accelerated outward from the center. At the distance $r_-$ the flow enters the interior of the white hole and leaves it at $r_+$.  The origin for this behavior is the gravitational repulsion caused by the negative gravitational mass placed in the center. It is compensated by the positive electromagnetic energy, so that the mass enclosed within the surface of the radius $r > r_0$ is positive. However, it becomes formally negative at $r < r_0$. Although the appearance of the negative mass is unphysical, formally one may solve the equations of motion for $r < r_0$, even though velocity $v$ is imaginary. The sketch of the structure of the charged non - rotated black - hole is represented in Fig. \ref{fig}. For the more detailed description of this space - time in the Painleve - Gullstrand reference frame see \cite{Hamilton:2004au}.

 It is widely believed \cite{Hamilton:2008zz}, that for the more realistic configurations, where space - time is not empty and contains matter, space - time metric behaves in a qualitatively different way at $r \sim r_0$. The negative mass and imaginary velocity $v$ do not appear. In the present paper we assume, that due to matter distributed somehow within Cauchy horizon the velocity field $|{\bf v}|$ inside the inner horizon is modified in such a way, that {\it it remains real and vanishes smoothly at $r=0$}.

\section{The "free falling vacuum", Hawking radiation, and Schwinger pair production}

\label{Sect_RE}

\begin{figure}
\begin{center}
 \epsfig{figure=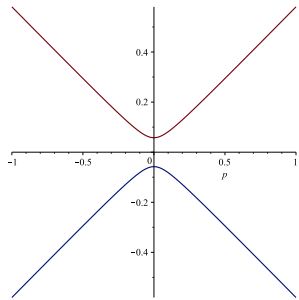,width=40mm,angle=0}
\end{center}
\caption{\label{fig.1}  The Fermi pockets are presented for the model with Dirac fermions with $m=0.1$ and $v=2$}
\end{figure}

In empty space everything is falling to the center of the black hole. For the simplicity we may imagine initial "vacuum" of the quantum field theory as a substance similar to the collection of the non - interacting particles. In this pattern it is falling with the velocity ${\bf v}({\bf r})$ defined above. In the small vicinity of any given point of space - time the vacuum of the field theory is to be considered at this stage as falling with constant velocity. In the accompanying reference frame the space - time is flat, and the Hamiltonians for the Weyl fermions are given by
\begin{equation}
H_0^\pm({\bf p}) =  \pm {\bf p} \sigma
\end{equation}
The mass term that mixes the left - handed and the right handed fermions remains unchanged. For the vanishing temperature the upper branch of  dispersion
$$ E(p) = \pm \sqrt{p^2 + m^2}$$
is empty while the lower branch is occupied.

It is well - known that the energy momentum tensor in the conventional quantum field theory is ultraviolet divergent. Therefore, this is difficult to speak of it on the basis of the known versions of the continuum field theory. The attempt to model vacuum as a substance described by a field has been made, for example, in the so - called q - theory described in \cite{Klinkhamer:2016jrt} and references therein. Since we do not know the correct description of vacuum as a substance this is difficult, however, to represent explicitly its equation of state. The above proposition adopted in the present paper models it as a collection of the non - interacting particles. This is the minimal, possibly, too primitive supposition. However, we suppose, that it reflects qualitatively the essence of the discussed phenomenon: the vacuum falls down in the same way as the non - interacting particles. It is expected, that the interactions (and their influence on the motion) may be considered as the perturbation. The Painleve - Gullstrand reference frame is useful for the description of this kind of motion because in this reference frame the boost performed in the small vicinity of any given point (with the velocity of the free falling) brings the metric to the (locally) flat form.

In the Painleve -- Gullstrand reference frame in a small vicinity of each point the dispersion of the fermions is given by
$$ E({\bf p}) = \pm \sqrt{p^2 + m^2} - {\bf v} {\bf p}$$
Between the two horizons the Dirac cone (appeared at $p \to \infty$) is overtilted. As a result, the piece of the occupied branch corresponds to the energies above zero. The corresponding fermions are able to escape from the black hole due to tunneling. This is the origin of Hawking radiation (see Fig. \ref{fig.0}). We give here its semiclassical description \cite{Volovik:1999fc,VolovikBH,Akhmedov:2006pg,Jannes:2011qp}.

For the particles moving near the external horizon the probability of tunneling is proportional to the exponent
$$
\Gamma \sim {\rm exp}\,\Big( - 2 \,{\rm Im} \int p_r dr  \Big)
$$
Here only the radial component of momentum $p_r$ is relevant. The  integral is to be taken over the contour in complex plane that connects the two disconnected pieces of the particle trajectory.

We obtain
\begin{equation}
\Gamma \sim {\rm exp}\,\Big( - 2 \, {\rm Im}\, \int \frac{|{\bf v}|{\cal E} + \sqrt{{\cal E}^2 - (m^2 + p_\perp^2)(1-{\bf v}^2)} }{1 - {\bf v}^2({\bf r})} dr \Big)
\end{equation}
which gives
\begin{equation}
\Gamma \sim {\rm exp}\,\Big( - \frac{\cal E}{T} \Big)
\end{equation}
with
$$
T = -\frac{1}{2\pi}\frac{d|{\bf v}|}{dr}\Big|_{r = r_+}
$$
Taking into account Fermi statistics we arrive at the distribution of the emitted fermions
$$
\rho_{\cal E} \sim \frac{1}{1 + e^{{\cal E}/T}}
$$
It is worth mentioning, that according to the original interpretation of the Hawking radiation \cite{Hawking:1974sw}  the particle is created directly outside of the horizon. The two interpretations give identical results. Thus following \cite{Volovik:1999fc,VolovikBH,Akhmedov:2006pg,Jannes:2011qp} we suppose, that they may be considered as the two different interpretations of the same phenomenon.


Let us consider for the completeness the process similar to Hawking radiation that occurs outside of the horizon - the Schwinger pair creation by electric field of the black hole.
In this case we consider the system outside of the horizon in a small vicinity of a certain  point, where the electric field may be considered as constant. We consider the system in the gauge, in which the Hamiltonian does not depend on time. For the particles moving outside of the horizon we have the following expression for the radial momentum:
\begin{equation}
p_r = \frac{|{\bf v}|W + \sqrt{W^2 - (m^2 + p_\perp^2)(1-{\bf v}^2)} }{1 - {\bf v}^2({\bf r})}
\end{equation}
Here
$$
W(r) = {\cal E} + E r
$$
and $E$ is the electric field. We should take $W < 0 $ at the initial position $r_i$ of the trajectory. At the final point $r_f = r_i + L$ the value of $W$ is to be positive. The transition between those two points corresponds to the transformation of the given occupied vacuum state to the excitation above vacuum. Between points $r_i$ and $r_f$ there is the classically forbidden region, where $W^2(r) - (m^2 + p_\perp^2)(1-{\bf v}^2)   <0$. At the turning points
$W = W^\pm =  \pm\sqrt{(m^2 + p_\perp^2)(1-{\bf v}^2)}$. Direct calculation gives the semiclassical tunneling exponent
\begin{eqnarray}
\Gamma &\sim & {\rm exp}\,\Big(2 i \int dr p_r \Big)\nonumber\\ & \sim &{\rm exp}\,\Big( - 2  \int_{W^-}^{W^{+}} \frac{dW}{E} \frac{\sqrt{-W^2 + (m^2 + p_\perp^2)(1-{\bf v}^2)} }{1 - {\bf v}^2} \Big)\nonumber\\ & = & e^{-\frac{\pi (m^2 + p_\perp^2)}{E}}
\end{eqnarray}
Integration over the initial momenta $p_\perp$ and $0< p_r < EL$ gives the rate of the pair production in the unit volume during the unit time (we should take into account, that the incoming flow has velocity $\partial_{p_r} {\cal E}(p)$):
 \begin{equation}
 \frac{d \Gamma}{dt} = \frac{E^2}{4 \pi^3 }{\rm exp}\Big(-\frac{\pi m^2 }{E}\Big)
 \end{equation}
which accidently coincides with the pair production rate in flat space - time in the presence of constant electric field \cite{Novikov:1980ni}. The similar result has been obtained also using another methods based on the Schwarzschild coordinates (see, for example, \cite{Khriplovich:1999gm}).  As a result the total pair production rate by the electric field of the black hole leads to the discharging of the black hole. This occurs  because one of the created particles in the pair (of the charge opposite to $Q$)  inevitably falls down to its interior. Thus we have
 \begin{equation}
 \frac{d Q}{dt}\Big|_{Schwinger} = -\frac{Q^2}{\pi^2 }\int_{r_+}^\infty \frac{dr}{r^2}{\rm exp}\Big(-\frac{\pi m^2 }{Q}r^2\Big)
 \end{equation}

\section{Vacuum reconstruction}

\label{True}

To be precise, the initial "vacuum" discussed in Sect. \ref{Sect_RE} is not the true vacuum. This state looks like a vacuum when one looks at it inside the small piece of the reference frame falling freely towards the center. If interactions are neglected, this state is composed of the occupied energy levels corresponding to the energies below zero in this accompanying reference frame. Globally the stationary accompanying reference frame does not exist. That's why the given state is not vacuum.

When interactions are taken into account, we should observe the transformation of the given state. During this transformation process the entropy is increased and reaches its maximum in the final equilibrium state. It is convenient to look at the final equilibrium state inside a stationary reference frame (i.e. the reference frame with the vierbein that does not depend on time). In this coordinate system the equilibrium state that gives maximal value of entropy corresponds to the generalized Gibbs distribution
\begin{equation}
\rho_{n} = {\rm exp}\, \Big(-\frac{E_n - \omega m_{\hat{\bf q}} }{T}\Big)\label{Go}
\end{equation}
where $E_{n}$ is the energy of the $ n $ - th state. Along with the energy the conserved additive quantity is angular momentum $L$ characterized by the certain values of total angular momentum $J^2 = l(l+1)$ and the certain value of its projection to the given axis $\hat {\bf q}$: $J_{\hat {\bf q}} = m_{\hat{\bf q}}$. Thus the equilibrium state is characterized by the four parameters: $\omega$, $\hat{\bf q}$, and $T$. The latter is identified with temperature, the first may be identified with the angular velocity $\omega$ of vacuum while $\hat{\bf q}$ is the rotation axis.
We may, in principle, choose any such reference frame. Our choice is the Painleve - Gullstrand coordinates. We take from the very beginning the spherically symmetric non - rotating free falling "vacuum". Therefore, the conserved angular momentum equals to zero, and we arrive at
\begin{equation}
\rho_{n} = {\rm exp}\, \Big(-\frac{E_n }{T}\Big)\label{G}
\end{equation}
instead of Eq. (\ref{Go}), which would be typical for the Kerr black hole solution. Vacuum reconstruction is accompanied by the Hawking radiation, which stops at the end of this process because the equilibrium is achieved. At the same time the discharging process due to the pair production by electric field may proceed if the electric charge is not evaporated completely during vacuum reconstruction.

\begin{figure}[!thb]
\begin{center}
{ \epsfig{figure=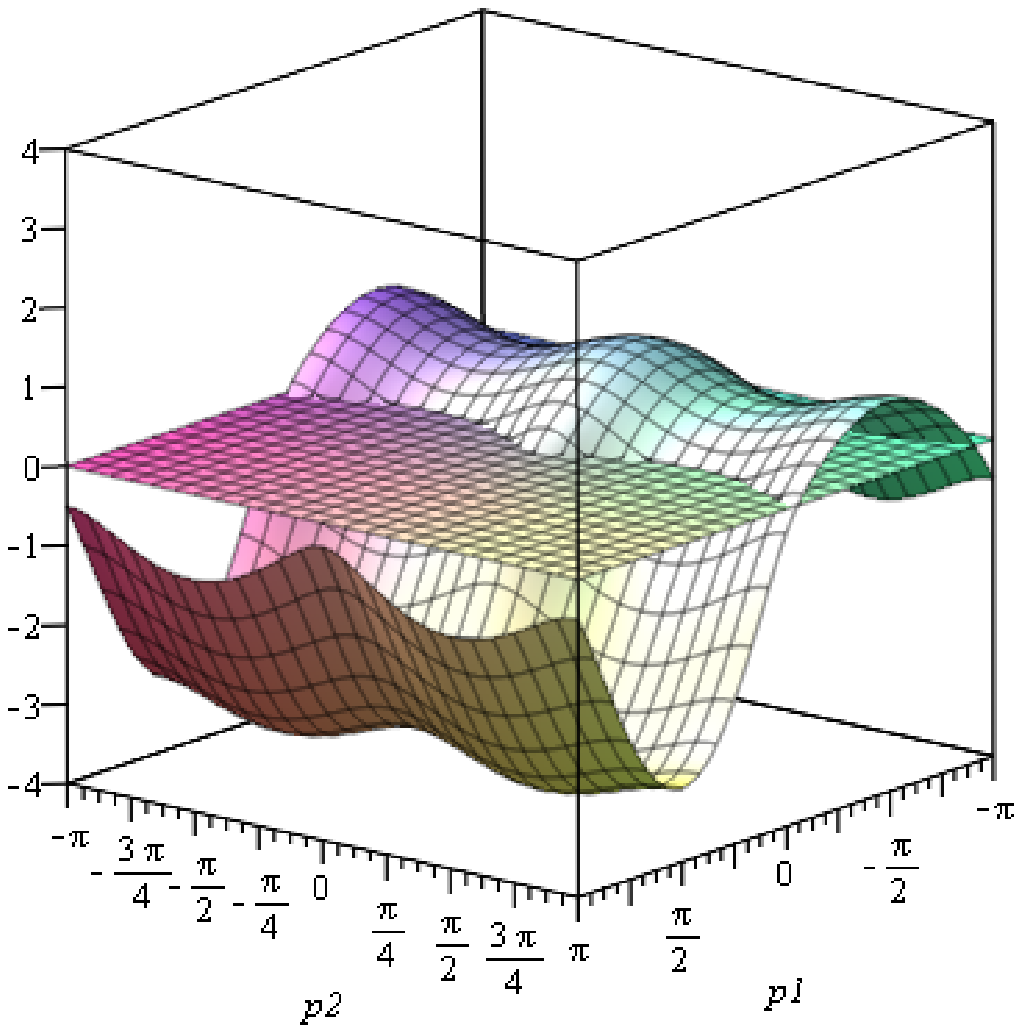,width=40mm,angle=0}}
{ \epsfig{figure=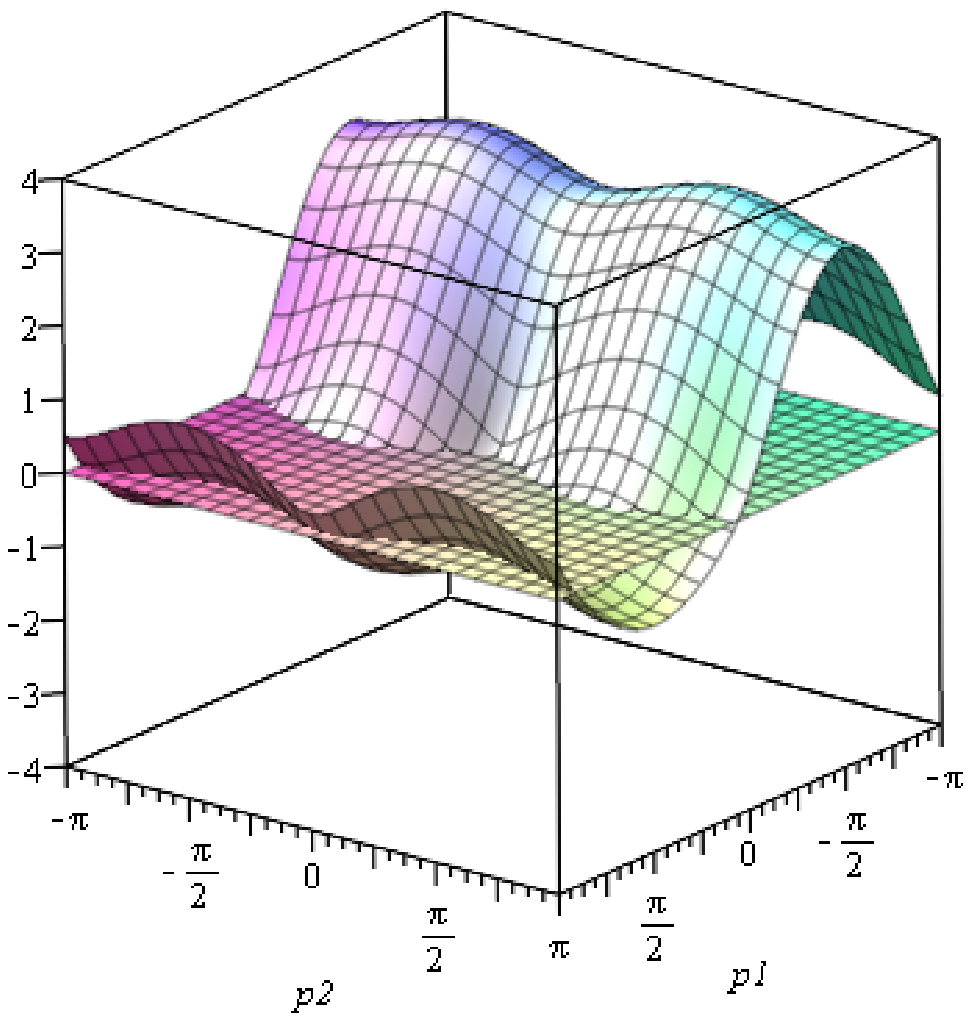,width=40mm,angle=0}}
\end{center}
\caption{\label{fig.23}  The two branches of spectrum for the model with lattice Dirac fermions with $m=0.1$, $r=0.1$, and $v=2$ is represented {\bf Left:} The lower branch of spectrum. Its intersection with plane ${\cal E}=0$ separates the hole Fermi pocket (of smaller area) from the area of occupied states.  {\bf Right:} The upper branch of spectrum. Its intersection with plane ${\cal E}=0$ separates the particle Fermi pocket (of smaller area) from the area of free states. }
\end{figure}

\begin{figure}
\begin{center}
\epsfig{figure=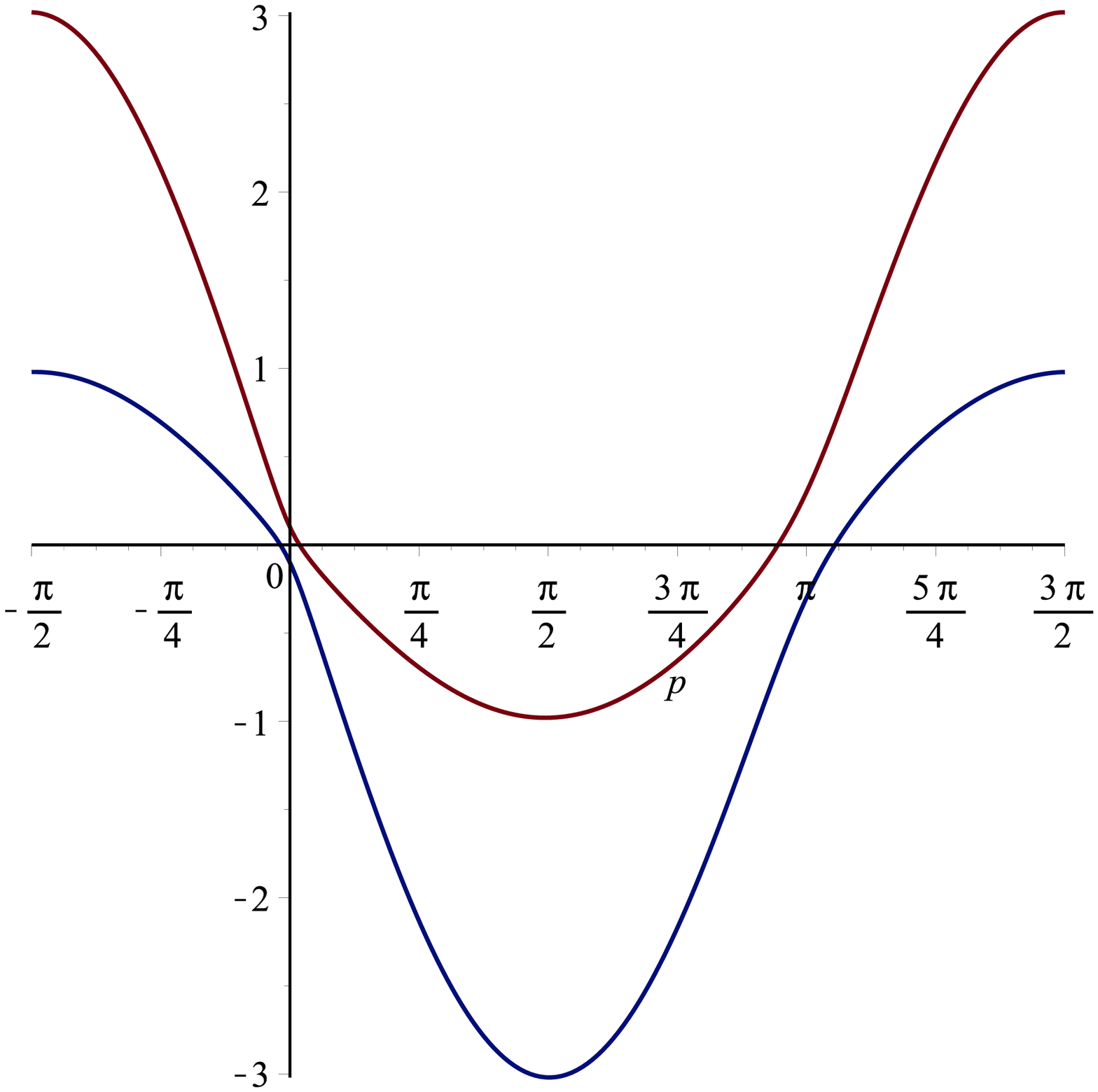,width=40mm,angle=0}
\epsfig{figure=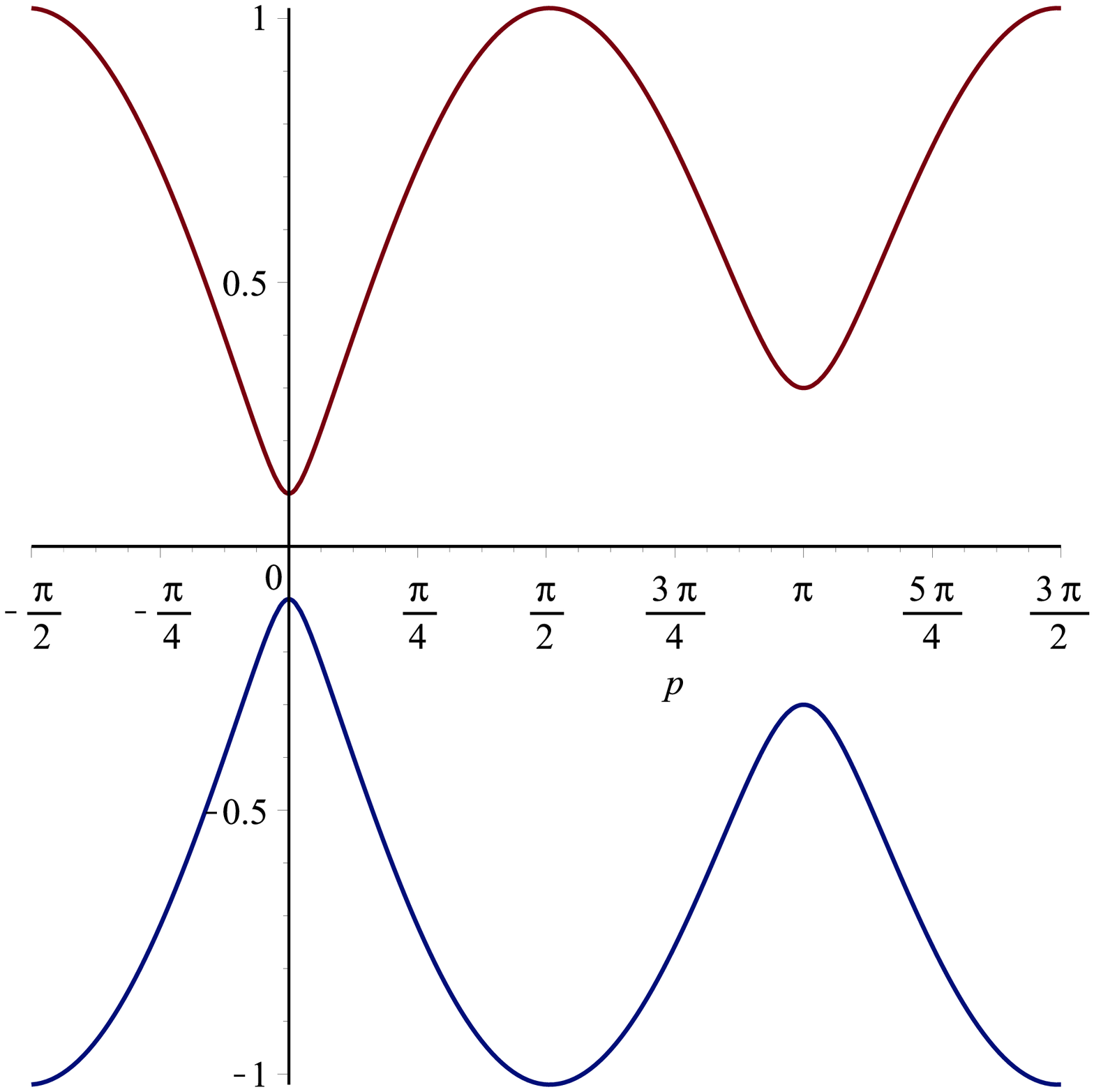,width=40mm,angle=0}
\end{center}
\caption{\label{fig.45}{\bf Left:}  The dispersion of the lattice Dirac fermions for $p_1,p_2 = 0$ is represented as a function of $p_3$ for $m=0.1$, $r=0.1$ and $v=2$. In the initial stage of the existence of the black hole the upper branch of spectrum symbolized by the upper red line is occupied, while the lower branch is not occupied.  When the equilibrium is achieved, the whole pattern is changed, and all states below zero are occupied in the true vacuum, while all states above zero are free. In both cases the occupied states of the left parts of the spectrum branches move towards the center of the black hole. That means that the corresponding constituent of vacuum is falling down towards the black hole interior. On the other hand the occupied states from the right hand sides of the branches are falling down from the interior towards the exterior of the black hole. For those states the black hole becomes the white hole. {\bf Right:} The dispersion of the lattice Dirac fermions for $p_1,p_2 = 0$ is represented as a function of $p_3$ for $m=0.1$, $r=0.1$ and $v=0$. This is the sketch of the situation that takes place outside of the horizon. There the massive fermion at $p_3=0$ corresponds to the ordinary matter while the fermion doubler at $p_3=\pi$ symbolizes extra massive matter that cannot be observed under normal conditions.}
\end{figure}

Even if originally the "temperature" was equal to zero, i.e. there were no particles above the free falling "vacuum", at the end of the process discussed here the temperature entering Eq. (\ref{G}) will in general be nonzero. At this stage if we neglect interaction, the system consists of the vacuum and excitations above it. Vacuum corresponds to the occupied states with negative energy. The excitations correspond to holes (the absence of  occupied states with negative energy) and particles (the occupied states with positive energy). We also make the further simplification considering the theory in a small vicinity of a given space - time point and neglecting the dependence of $\bf v$ on $r$.

\begin{figure}
\begin{center}
\epsfig{figure=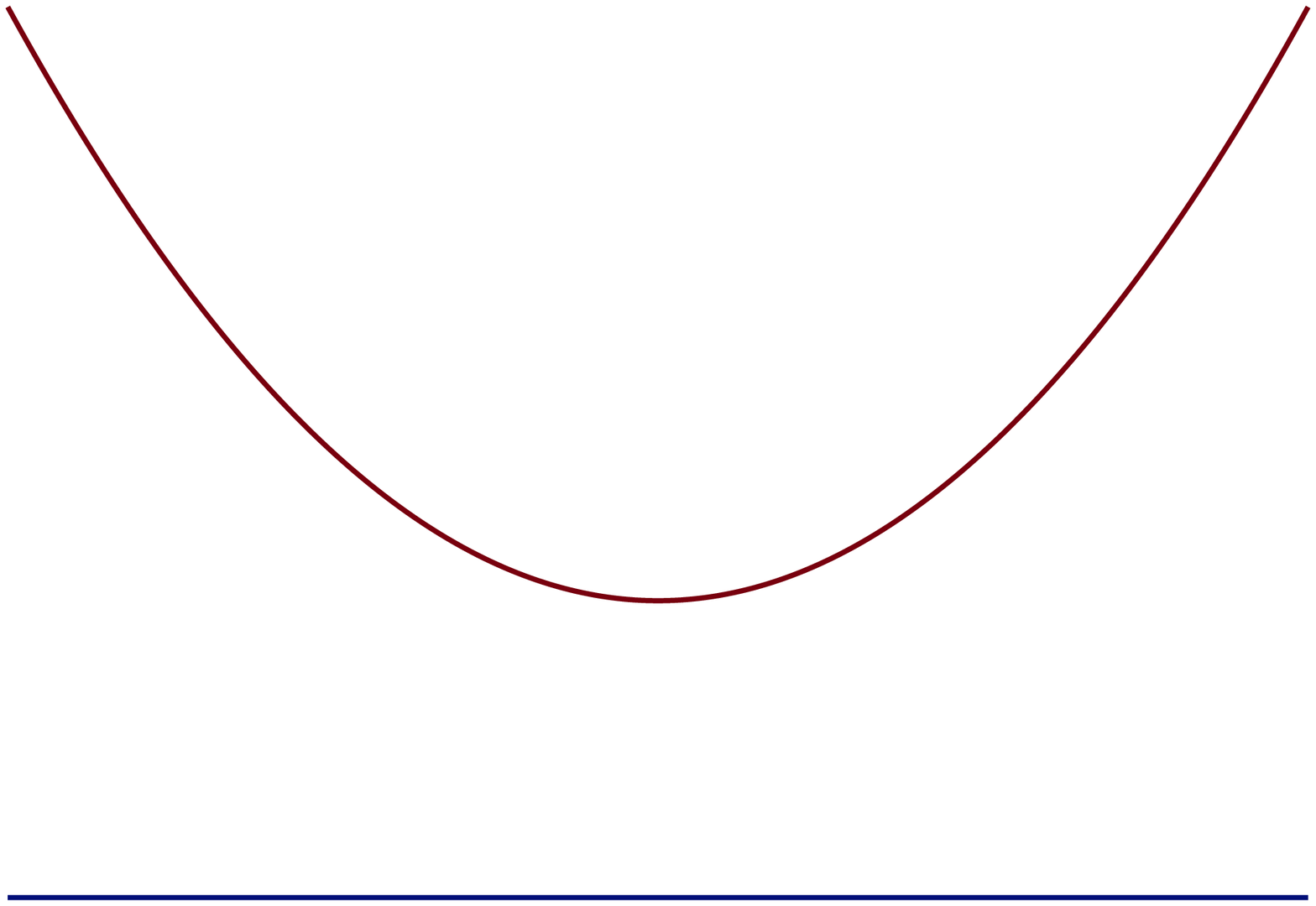,width=20mm,angle=0}
\epsfig{figure=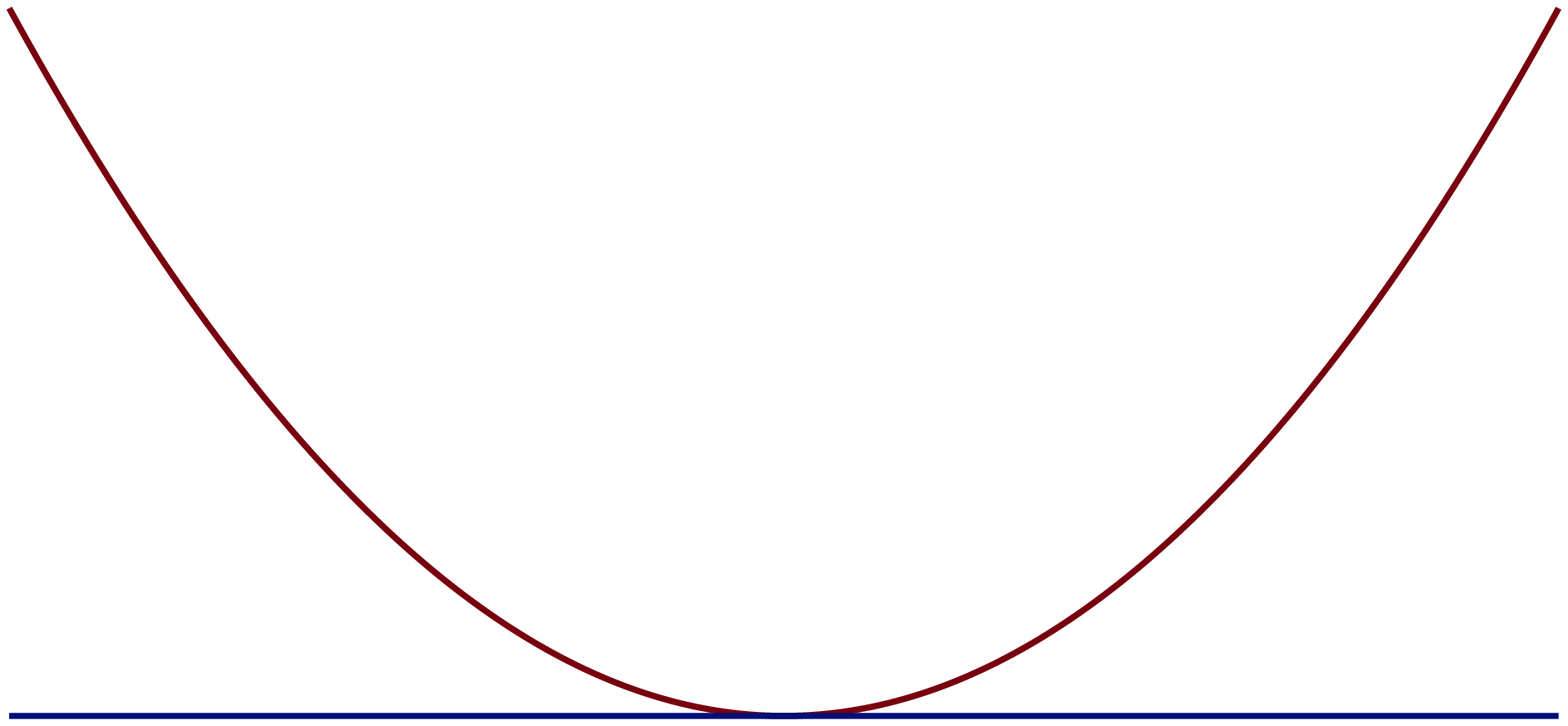,width=20mm,angle=0}
\epsfig{figure=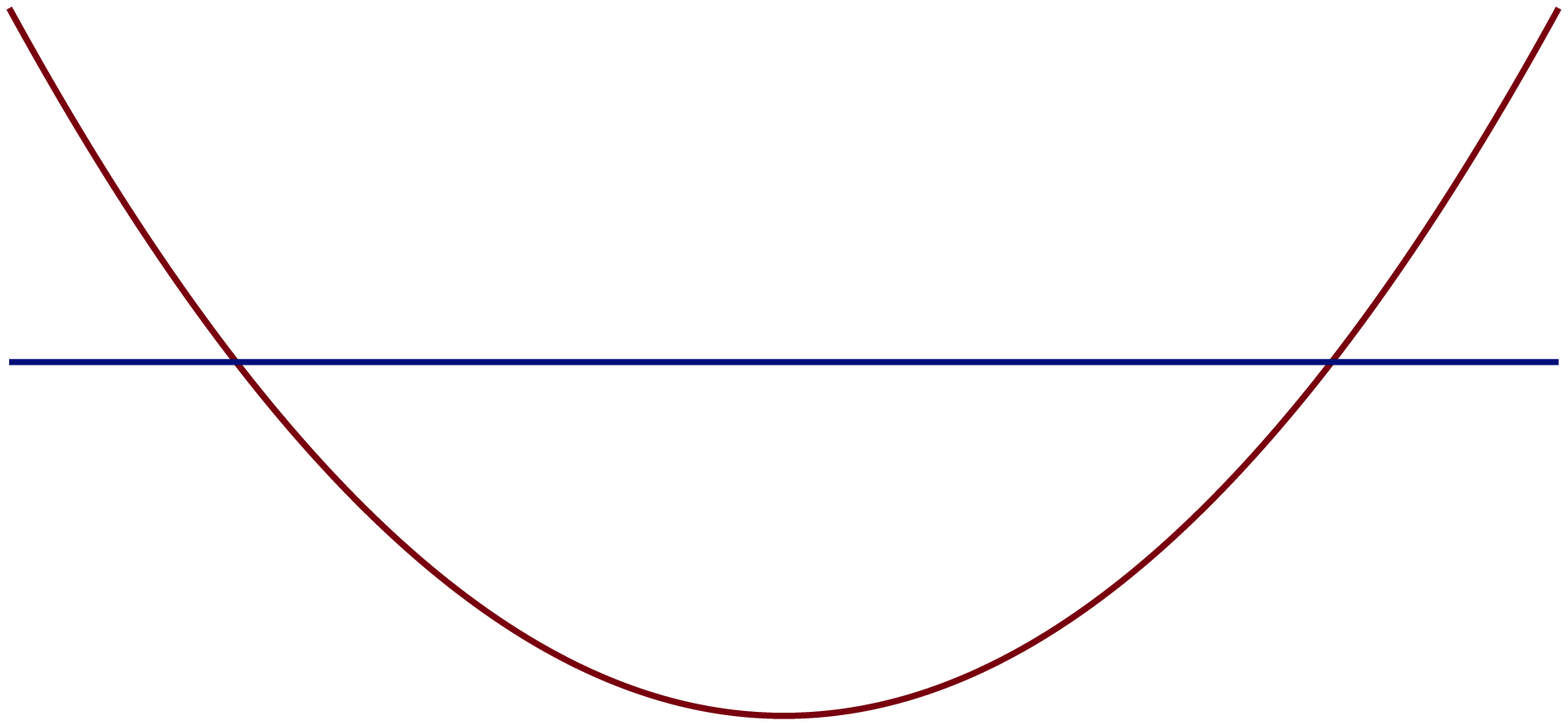,width=20mm,angle=0}
\end{center}
\caption{\label{fig.67} Sketch of the transformation of the upper branch of spectrum while moving towards the black hole. {\bf Left:} outside of the horizon. {\bf Center:} At the horizon. {\bf Right:} Inside the horizon. }
\end{figure}

Outside of the external horizon we have the ordinary Weyl fermions with  mass terms, i.e. the Dirac fermions. The Dirac cones are tilted, but not overtilted. Therefore, the occupied vacuum states are separated from the region of particles by the energy gap.

Inside the horizon (more precisely - for $r_- < r < r_+$) the energy dispersion is
$$ E({\bf p}) = \pm \sqrt{p^2 + m^2} - {\bf v} {\bf p}$$ with $|{\bf v}|>1$. The would be Dirac cone is overtilted. Therefore, the part of the upper branch of the spectrum is occupied (that is not occupied when the cone is not overtilted). At the same time the piece of the lower branch of the spectrum becomes free of particles. This situation is similar to that of the type II Weyl fermions that were discovered recently in the so - called type II Weyl semimetals. Instead of the energy gap that separates the occupied and non  - occupied branches of spectrum, the Fermi surface appears. In the vicinity of the would be Fermi point $P^{(0)} = 0$ there are the two pieces of the Fermi surface that form the particle Fermi pocket and the hole Fermi pocket correspondingly. For the description of the notion of the Fermi pockets we refer to \cite{W2}. The Fermi pocket is the part of the Fermi surface, that is separated from the other parts of the Fermi surface. In the case of the conventional type II Weyl fermions the two Fermi pockets are compact and touch each other at the type II Fermi point. One of the pockets encloses the region of occupied states (the particle Fermi pocket) while the other enclosed the region of empty states (called the hole Fermi pocket). In our case the Fermi pockets do not touch each other, there is a small distance between them because of the mass term for the fermions. In the toy model presented below the Fermi pockets are compact as in the condensed matter systems. In the present section that does not rely on the UV completion of the Standard Model they remain infinite.

In Fig \ref{fig.1} we represent the form of the Fermi pockets for the value of the mass $m = 0.1$ measured in certain energy units, and the value of velocity $v = 2$. The important property of the system clearly demonstrated by this figure is that the Fermi surface (here its slice for $p_3 =0$ in the plane ($p_1,p_2$) is represented) is prolonged to the infinite values of momenta. The low energy physics is located in the vicinity of the Fermi surface. Therefore, we already cannot restrict ourselves by the values of momenta close to $P^{(0)} = 0$, where the Standard model of the fundamental interactions is relevant. Now the low energy physics appears also in the region of large momenta, where the Standard Model does not work. We inevitably come to the necessity to consider an ultraviolet completion of the Standard Model.

Unfortunately this is difficult or even impossible to estimate the time of the relaxation to equilibrium. In order to do this we not only need to take into account the interactions. We should consider the whole model including its excitations that reside far from the point in momentum space $P^{(0)}=0$, where the Standard Model resides. This is because in equilibrium the dispersion vanishes far from $P^{(0)}=0$, and, therefore, the corresponding excitations are relevant. If the reader would rely on the pattern that is based on the Standard Model only and does not take into account its ultraviolet completion, then the particles live between the two horizons during the limited time period  around $r_+$. The Standard Model particles between the two horizons move only towards the center. However, in the toy model considered below in Sect. \ref{Sect_EQ} there are the excitations that are able to move in the opposite direction. Furthermore, in the idealized model with the empty charged black hole spacetime the Standard Model particles cross the Cauchy horizon, proceed moving at  $r< r_-$, and, although in this region they may move backward as well, they are able to drop to the analytically continued space - time, enter the white hole, and then, the new Universe. As it was mentioned above, in the present paper we assume, that the structure of space - time at $r < r_-$ differs from this unrealistic pattern. Instead, we suppose, that velocity $v$ remains real for $0 < r < r_0$ and vanishes at $r = 0$ only. We also imply that there is no analytical continuation of spacetime. That means, that matter is distributed inside the Cauchy horizon in a nontrivial way (see, for example, \cite{Hamilton:2008zz}).  The Standard Model particles that fall to this region of space - time do not disappear in the new Universe. Instead, they may be transformed due to interactions to the excitations of the UV completion of the Standard Model. The latter are able to move towards the Cauchy horizon, cross it and move further, even outside of the outer horizon. The mentioned transformations of particles, and the mentioned backward motion, as well as the interactions in the UV completion of the Standard Model, drive the relaxation of the system to equilibrium. However, the further elaboration of this issue is out of the scope of the present paper.


\section{The toy model inspired by the type II Weyl semimetals}

\label{Sect_EQ}

The use of lattice regularization at the present moment is the only rigorous way to define the quantum field theory. In any lattice model the Fermi surface is closed in compact momentum space. The region of momentum space distant from the point $P^{(0)}$ is irrelevant for the description of physics at low energies outside of the horizon. However, this region becomes relevant inside the horizon for the description of the low energy physics. We illustrate this simple observation by the lattice model with the Hamiltonian
\begin{eqnarray}
H &=& \sum_{k = 1,2,3} {\rm sin}\,p_k \gamma^0\gamma^k - (m + r \sum_{k = 1,2,3}(1-{\rm cos} \,p_k))\gamma^0 \nonumber\\ && - v \, {\rm sin}\, p_3\label{HL}
\end{eqnarray}
Here $\gamma^k$ for $k = 0,1,2,3$ are the Dirac matrices in chiral representation:
\begin{eqnarray}
\gamma^{i}=\begin{pmatrix}
0 &\;\sigma^{i} \\
-\;\sigma^{i} & 0 \\

\end{pmatrix} ,& \gamma^{0}=\begin{pmatrix} 0 & 1 \\
1 & 0
\end{pmatrix}
\end{eqnarray}
where $\sigma^{i},i=1,2,3$ are the Pauli matrices. It is worth mentioning that in this model the local Lorentz symmetry is broken explicitly. It is restored only in the vicinity of the point $P^{(0)} = 0$, where the given Hamiltonian receives the form
\begin{eqnarray}
H \approx \begin{pmatrix}
{\bf p} \sigma - {\bf p} {\bf v}& m \\
m & -{\bf p} \sigma - {\bf p} {\bf v}  \\
\end{pmatrix}
\end{eqnarray}
Here ${\bf v} = (0,0,v)$. In order for the local Lorentz symmetry to be restored we need $m \ll 1$ in the lattice units. That means $m = m_{phys}a$, where $a$ is the lattice spacing while $m_{phys}$ is the mass of the excitation living near $P^{(0)}$ measured in physical units. The energy scale of the violation of Lorentz symmetry is given by $\Lambda = \pi/a$. At the energies much smaller than $\Lambda$ for $v < 1$ the relevant excitations are those massive fermions living near $P^{(0)}$. The typical values of physical momenta $p^i_{phys} = p^i \frac{1}{a}$ are also much smaller than $\Lambda$. The excitations with $|{\bf p}_{phys}| \sim \Lambda$ are suppressed at $r \sim 1$. In particular,
close to the point $P^{(1)} = (0,0,\pi)$ we have
\begin{eqnarray}
H &\approx & \begin{pmatrix}
\delta{\bf p} \sigma^\prime + \delta{\bf p} {\bf v}& (m + 2 r) \\
(m + 2 r) & -\delta{\bf p} \sigma^\prime + \delta{\bf p} {\bf v}  \\
\end{pmatrix},\\ && \delta {\bf p} = {\bf p} - (0,0,\pi)\nonumber
\end{eqnarray}
with $\sigma^\prime_1 = \sigma_1$, $\sigma^\prime_2 = \sigma_2$, $\sigma^\prime_3 = -\sigma_3$. The mass of the corresponding excitations written in physical units is $m^\prime_{phys} \sim 2 r \frac{\Lambda}{\pi}$. At small enough energies those excitations cannot be seen. It is worth mentioning, that at the present moment the scale of the violation of the local Lorentz symmetry remains model - dependent. As an example we refer to the lower bound on the Lorentz scale violation in the theories with the Lifshitz - Horava gravity \cite{Lifshitz}: $\Lambda \ge 10^{16} $GeV. In certain schemes of the UV completion of the SM there may even exist the lower bounds on the Lorentz symmetry violation \cite{Liberati}.

The situation is changed drastically at $v > 1$. Namely, the model with the Hamiltonian of Eq. (\ref{HL}) reveals an analogy with the  type II Weyl semimetals. Its main feature is that at $v>1$ the Dirac cone is overtilted. However, there is the difference: instead of the two Weyl points of opposite chiralities, separated by the distance in momentum space, Eq. (\ref{HL}) describes the Dirac points separated in momentum space (in the limit $m\to 0$). Each Dirac point corresponds to the pair of the left - handed and the right - handed fermions. Therefore, at $m=0$ Eq. (\ref{HL}) may be considered as the modeling Hamiltonian for a  material (not yet discovered experimentally) that could be called the type II Dirac semimetal. Moreover, since $m \ne 0$, at $v <1$ we deal with the insulator instead of the semimetal while at $v>1$ we deal with a metal with the Fermi surface of very specific form, which is reduced to the Fermi pockets touching each other in the limit $m\to 0$.

We represent in Fig. \ref{fig.23} the intersection of the two branches of spectrum with plane ${\cal E} = 0$ for $m = 0.1$, $r=0.1$, and $v  = 2$. One can see, that the two Fermi surfaces are closed and their pieces are situated close to the point $P^{(0)}$ as well as close to the positions of the fermion doublers. The latter are extra heavy fermions, which disappear from the low energy effective theory outside of the horizon (see Fig. \ref{fig.45}, right hand side). Inside the horizon those regions of momentum space are relevant for the low energy description.



The above consideration means that in addition to ordinary fermionic matter outside of the horizon in a certain vicinity of it there may exist the additional matter, which cannot manifest itself at low energies. For the lattice model with the Hamiltonian of Eq. (\ref{HL}) the dispersion is represented in Fig. \ref{fig.45}, right hand side. There is the additional massive fermion at $p_3 = \pi$. This additional fermion is more massive than the one placed at $p_3 = 0$. Therefore, it decouples at low energies. (In this figure the mass is not essentially larger than the mass of the usual fermion, but in realistic theory it might be many orders of magnitude larger. It may actually be estimated as $m^\prime_{phys} \sim 2 r \frac{\Lambda}{\pi}$, where $\Lambda$ is the scale of the violation of local Lorentz symmetry.) It is interesting, that with respect to these extra massive fermions the black hole plays the role of the white hole. This is illustrated by Fig. \ref{fig.45}, left hand side,
where the behavior of dispersion inside the black hole in the Painleve - Gullstrand reference frame is represented schematically. One can see, that the slope of the two lines near to $p_3 = \pi$ is opposite to that of the lines at $p_3=0$.  This is why the black hole with respect to these additional excitations plays the role of the white hole.  The essence of the proof is that {\it the velocity of the quasiparticles is given by the derivative of the dispersion} (of the dependence of energy on momentum). One can see, that the slope of the dispersion around $P^{(0)}=0$ corresponds to velocity directed towards the center of the black hole, which means that the particles described by the Standard Model move towards the center of the black hole only. However, far from the point $P^{(0)} = 0$ {\it the slope inevitably changes its direction}. As a result the corresponding excitations move towards the exterior of the black hole. And in the considered toy model they cannot move towards the center. That actually means, that with respect to them the black hole is nothing but the white hole.

After the vacuum reconstruction the thermal quasiparticles exist in the vicinity of the piece of the Fermi surface that is formed near to the position (in momentum space) of the mentioned extra massive additional matter. Those quasiparticles are able to traverse the horizon towards the exterior of the black hole. If their energy $\cal E$ is smaller, than their mass at infinity, then they cannot reach it. The corresponding turning point $r^\prime$ is the solution of equation
\begin{equation}
{\cal E} = m^\prime \sqrt{1 - {\bf v}^2(r^\prime)}
\end{equation}
Here $m^\prime$ is the mass of these extra excitations (in the toy model presented above $m^\prime = m + 2 r$).  This additional matter under usual conditions cannot be observed. However, it escapes from the black hole, and in a certain vicinity of the horizon (outside of it) it may have the sufficiently small energy. There this kind of matter participates in the low energy processes. We conclude, that the excitations, which under normal conditions may appear in the processes with the unreachable energy, may be observed near the horizon. There those excitations participate in the low energy processes, may be transformed to the ordinary Standard Model particles, and thus may be indirectly detected somehow by distant observers.

Let us add a remark on the distribution of electric charge inside the black hole. In equilibrium theory the system with the Fermi surface behaves like a metal. Therefore, inside the black hole between the two horizons the vacuum behaves like metal. As a result, the electric charge $Q$ placed inside the second (Cauchy) horizon induces the appearance of charge $-Q$ at the inner horizon and charge $+Q$ at the outer horizon (if space for $r < r_-$ is insulating). If space for  $r < r_-$ would be conducting,  the electric charge remains at the outer horizon only. The latter case is realized if in this region the geometry differs somehow from the standard Reissner-Nordstrom one. This, in turn, may occur if the extra gravitating matter is present there. In any case the electric field is absent between the two horizons. It is interesting, that the states from the occupied vacuum branches near the point $P^{(0)}=0$ always give the contribution to electric current directed towards the center of the black hole. The total current of vacuum is vanishing according to the obvious identity $\int d^3p \partial {\cal E} = 0$. Therefore, the compensating electric current is carried by the region of momentum space far from  $P^{(0)}=0$. And this is due to the corresponding states the electric charge is accumulated at the outer horizon.
The above pattern is perturbed due to the pair production by the electric field of the black hole outside of the horizon. The description of this process was given briefly in Sect. \ref{Sect_RE}. The particles/holes of charge opposite to $Q$ traverse the horizon and lead to the discharge of the black hole.

In the above considerations we omitted almost completely the bosonic degrees of freedom. Below we discuss briefly the extension of the above fermionic model of Eq. (\ref{HL}) to the case of the charged scalar bosons $\Phi$. The consideration of the gauge field and the gravitational fields (the vierbein and the spin connection) is similar. First of all, let us write down the corresponding action that is to be used instead of the Hamiltonian of Eq. (\ref{HL}): \begin{eqnarray}
  S&=& \int_{-\pi}^{+\pi} d^3 k \int_{-\infty}^{+\infty}d\omega  \Phi^+(k,\omega) \Big(\omega^2 (1-v^2)\nonumber\\&&- 2 \sum_{k = 1,2,3} (1-{\rm cos}\,(p_k )) - 2 v {\rm sin} (p_3) \omega - m^2 \Big)\Phi(k,\omega) \label{HL2}
\end{eqnarray}
This action is written in momentum space for simplicity. By $\omega$ we denote the frequency of the field $\Phi$ that may be also understood as the energy of the corresponding excitations.  The dispersion of the scalar quasiparticles has the form:
$$
\omega(k) = \frac{1}{1-v^2}\Big(\pm \sqrt{\Big(m^2 + 2\sum_{k = 1,2,3} (1-{\rm cos}\,(p_k ))\Big)(1-v^2)+ v^2 {\rm sin}^2 (p_3)} + v {\rm sin} (p_3)  \Big)
$$
At $v = 0$ and $m=m_{phys}a,p_k = p_{k,phys}a \to 0 $ we recover the dispersion of the relativistic scalar bosons in flat space - time corresponding to the mass and momentum in physical units $m_{phys}$, $p_{phys}$. At nonzero $v$ this dispersion is changed in accordance to the continuum Klein - Gordon equation written in the background of the black hole in the Painleve - Gullstrand reference frame. For $v < 1 $  at the values of $p$ far from zero the energy remains large, which means that the corresponding excitations are not relevant for the low energy physics. Inside the black hole horizon, when $v > 1$, the dispersion of the scalar excitations differs essentially from the above mentioned form. First of all, the imaginary part of energy appears at certain values of momenta signalizing that the corresponding excitations are not stable. Next,  the surface in momentum space appears, where the real part of the energy of the scalar excitations is equal to zero.

\section{Conclusions}

\label{Sect_CON}

We discussed the quantum field theory in the presence of the background charged non - rotating black hole. In our approach to the quantum field theory we understand it as a theory similar to a certain extent to the electronic theory of solids. This point of view is equivalent to the consideration of the lattice regularization for the quantum field theory as the true theory, to which the known continuum field theory is only a low energy approximation. (Recall, that the lattice regularization is the only rigorous way to define the quantum field theory non - perturbatively.) In this approach the invariance under general coordinate transformations and the gravitational fields might be the emergent phenomena. However, the precise way how they enter the unified theory is not important for us. From this unknown theory we only require that momentum space is compact.

It is also not important, what occurs behind the Cauchy horizon. Actually, we need the conventional form of the black hole solution only outside of the outer horizon and just behind it. For the definiteness we assume, that the conventional black hole solution remains valid until the Cauchy horizon, while behind it metric is modified in such a way, that velocity $|{\bf v}|$ (entering expression for the metric in Painleve - Gullstrand reference frame) remains real and vanishes smoothly at $r=0$. This modification is implied to be the consequence of the processes that are not described by classical gravitational theory in accordance to the supposition of \cite{Hamilton:2008zz}. This supposition is justified at least if
$r_0 = \frac{Q^2}{2 M} \sim \frac{1}{m_p}$, so that the quantum gravity is to be taken into account at these distances.

We assume, that according to the scenario of \cite{VolovikBH} the vacuum of the black hole is reconstructed: from the conventional state the black hole comes to the new one with the essentially different properties. This state is considered in the Painleve - Gullstrand reference frame. We argue, that in the consistent quantum field theory (understood in the mentioned above way), which describes this equilibrium state, the new type of matter exists. A part of this matter has the properties that are in a certain sense opposite to those of the ordinary matter described by the Standard Model. In order to illustrate this we consider the toy model inspired by the type II Weyl semimetals.

In this model the Standard Model particles are modeled by the excitations living near to the point $P^{(0)} = 0$ in momentum space. In addition, this model includes the extra massive particles that appear in the vicinities of several points in momentum space that are distant from $P^{(0)}$. Those extra particles actually are the incarnations of the so - called fermion doublers, which are well - known in the framework of the lattice field theory. Under normal conditions the mentioned extra particles cannot be observed because they are extra massive.

After vacuum reconstruction the new vacuum is formed. Both ordinary and additional matter form the Fermi surface. As a result both types of excitations appear at low energies. The peculiar feature of the considered toy model is that only the ordinary matter (that resides near the point $P^{(0)}$ in momentum space) experiences gravity in a conventional way. The gravitational force acts on the mentioned above extra massive particles in the direction opposite to that of the force acting on the SM particles. As a result
the additional extra massive matter is able to escape from the interior of the black hole directly, not through the tunneling mechanism of Hawking radiation. The corresponding particles with small initial energy cannot reach the infinitely distant observer, but they are able to exist outside of the horizon in a certain vicinity of it.  Among the other consequences the ability of the extra massive matter to escape from the black hole may resolve the so - called information paradox \cite{Mann,information,BHI}.
 We argue, that the same pattern is likely to take place in a variety of the self - consistent quantum field theories that attempt to describe matter in the presence of the black holes.

In the more realistic theory the pattern of the energy branches is more complicated. It may differ somehow from the pattern discussed above. Let us  assume, that inside the black hole the Fermi surface is closed and finite while the system far from the black hole is gapped. Also let us assume that the energy levels are transformed smoothly while moving towards the horizon. At a certain moment the upper branch touches the plane ${\cal E} = 0$. Next, it starts crossing the plane. In the non - marginal case the crossing occurs as shown in Fig. \ref{fig.67}. The position of the horizon corresponds to the moment of touching. Inside the horizon the pattern is as represented at the right panel of Fig. \ref{fig.67}. One can see, that at the crossing points the slopes in this non - marginal case have opposite signs.  Let us assume in addition the symmetry between particles and antiparticles, i.e. the symmetry between the particles and holes. Then, the lower branch of spectrum is transformed in the similar way to the upper branch. It crosses the plane ${\cal E}=0$ from below at the same moment, when the upper branch crosses this plane from above. Altogether, just after the moment of touching (i.e. just behind the horizon) we have the same pattern as above for the model with the lattice Dirac fermions with the Hamiltonian of Eq. (\ref{HL}). There the excitations exist near the Fermi surface, such that for them the black hole plays the role of the white hole. If the particle - hole symmetry is broken, then the pattern still remains similar. However, the Fermi surfaces made by the upper and the lower branches might appear non - simultaneously. While moving further inside the horizon, the slopes of the energy branches at the Fermi surface may, in principle, acquire the same sign. But this does not change our conclusion about the existence of the unusual low energy excitations just behind the horizon.

We conclude the following.
\begin{enumerate}
\item{} If the black hole undergoes the transition to the equilibrium state proposed in \cite{VolovikBH}, then inside such a black hole the Standard Model cannot describe the low energy physics. This occurs because the states are involved in dynamics that correspond to the region of momentum space, where the Standard Model already does not work and is to be replaced by its ultraviolet completion. We thus would be able to observe this ultraviolet completion if its excitations are able to escape from the black hole and transfer the information about the interior of the black hole to the distant observer.

\item{} In any self - consistent quantum field theory in the presence of the equilibrium black hole there are the excitations that may escape from the black hole (directly, not through the Hawking mechanism) if the theory obeys the following two conditions:
    1) The Fermi surface within the horizon is closed and finite.  2) While traversing the horizon the energy levels are transformed smoothly.

    The second condition matches the common lore, according to which the
physical quantities are not singular at the horizon in the reference frame accompanying
the falling observer. The first condition also looks rather natural. At zero temperature the number of states within the black hole is large, but it cannot be infinite. Those states are enclosed by the Fermi surface, and, therefore,  the latter also cannot be infinite. This occurs, in particular, if the given theory is defined in lattice regularization with the finite ultraviolet cutoff as in the proposed above toy model.

    The particles, which escape from the black hole, may exist in the small region of space outside of the horizon. Within this region they interact with the Standard Model particles, which, in turn, are able to reach the distant observer.

\item{} There is only one alternative to the mentioned above two points: the equilibrium state of the black hole is never achieved. The black hole in this case cannot be described by the equilibrium quantum field theory. Its evolution is governed by the kinetic theory. Due to interactions the particles leave the occupied states above zero (the left part of Figure \ref{fig.0}, the lower left branch) and fall down to the vacant states below zero (the right upper branch). This is the vacuum reconstruction, which never achieves its goal.

\end{enumerate}

The author kindly acknowledges useful discussions with G.E. Volovik, I. Ben - Dayan and B.Rosenstein.


\begin{thebibliography}{99}

\bibitem{VolovikBH}
  P.~Huhtala and G.~E.~Volovik,
  ``Fermionic microstates within Painleve-Gullstrand black hole,''
  J.\ Exp.\ Theor.\ Phys.\  {\bf 94} (2002) no.5,  853
   [Zh.\ Eksp.\ Teor.\ Fiz.\  {\bf 121} (2002) no.5,  995]
  doi:10.1134/1.1484981
  [gr-qc/0111055].

\bibitem{Schwarzschild}  Schwarzschild, K. (1916). "Uber das Gravitationsfeld eines Massenpunktes nach der Einsteinschen Theorie". Sitzungsberichte der Koniglich Preussischen Akademie der Wissenschaften. 7: 189–196. and Schwarzschild, K. (1916). "Uber das Gravitationsfeld einer Kugel aus inkompressibler Flussigkeit nach der Einsteinschen Theorie". Sitzungsberichte der Koniglich Preussischen Akademie der Wissenschaften. 18: 424–434.




\bibitem{Gullstrand}
Allvar Gullstrand, “Allgemeine L¨osung des statischen
Eink¨orperproblems in der Einsteinschen Gravitationstheorie”,
Arkiv. Mat. Astron. Fys. 16(8), 1–15 (1922)

\bibitem{Painleve}
Paul Painleve, “La mecanique classique et la theorie de la
relativite”, C. R. Acad. Sci. (Paris) 173, 677–680 (1921).

\bibitem{Hamilton:2004au}
  A.~J.~S.~Hamilton and J.~P.~Lisle,
  ``The River model of black holes,''
  Am.\ J.\ Phys.\  {\bf 76} (2008) 519
  doi:10.1119/1.2830526
  [gr-qc/0411060].



\bibitem{Doran:1999gb}
  C.~Doran,
  ``A New form of the Kerr solution,''
  Phys.\ Rev.\ D {\bf 61} (2000) 067503
  doi:10.1103/PhysRevD.61.067503
  [gr-qc/9910099].

\bibitem{Volovik:1999fc}
  G.~E.~Volovik,
  ``Simulation of Painleve-Gullstrand black hole in thin He-3 - A film,''
  JETP Lett.\  {\bf 69} (1999) 705
   [Pisma Zh.\ Eksp.\ Teor.\ Fiz.\  {\bf 69} (1999) 662]
  doi:10.1134/1.568079
  [gr-qc/9901077].

\bibitem{Hawking:1974sw}
  S.~W.~Hawking,
  ``Particle Creation by Black Holes,''
  Commun.\ Math.\ Phys.\  {\bf 43} (1975) 199
   Erratum: [Commun.\ Math.\ Phys.\  {\bf 46} (1976) 206].
  doi:10.1007/BF02345020

\bibitem{Parikh:1999mf}
  M.~K.~Parikh and F.~Wilczek,
  ``Hawking radiation as tunneling,''
  Phys.\ Rev.\ Lett.\  {\bf 85} (2000) 5042
  doi:10.1103/PhysRevLett.85.5042
  [hep-th/9907001].

\bibitem{Akhmedov:2006pg}
  E.~T.~Akhmedov, V.~Akhmedova and D.~Singleton,
  ``Hawking temperature in the tunneling picture,''
  Phys.\ Lett.\ B {\bf 642} (2006) 124
  doi:10.1016/j.physletb.2006.09.028
  [hep-th/0608098].





\bibitem{Jannes:2011qp}
  G.~Jannes,
  ``Hawking radiation of $E < m$ massive particles in the tunneling formalism,''
  JETP Lett.\  {\bf 94} (2011) 18
  doi:10.1134/S0021364011130091
  [arXiv:1105.1656 [gr-qc]].

\bibitem{Volovik2003}
G.E. Volovik,
{\it The Universe in a Helium Droplet},
Clarendon Press,  Oxford (2003)

\bibitem{liquid}
W. G. Unruh, Phys. Rev. Lett. 46, 1351 (1981)

M. Visser, ‘‘Acoustic black holes,’’ http://xxx.lanl.gov/abs/gr-qc/9901047,  Proceedings of the
1998 Peniscola Summer School on Particle Physics and Cosmology, Springer-Verlag.

T. A. Jacobson and G. E. Volovik, Phys. Rev. D 58, 064021 (1998); JETP Lett. 68, 874 (1998)

Visser, Class. Quantum Grav. 15, 1767 (1998)

\bibitem{semimetal_effects6}
S. Parameswaran, T. Grover, D. Abanin, D. Pesin, and A. Vishwanath,
{\sl ``Probing the chiral anomaly with nonlocal transport in Weyl semimetals},
Phys. Rev. X {\bf 4}, 031035 (2014) [arXiv:1306.1234].




\bibitem{semimetal_effects10}
M. Vazifeh and M. Franz,
{\sl ``Electromagnetic response of weyl semimetals''},
Phys. Rev. Lett. {\bf 111}, 027201 (2013) [arXiv:1303.5784].

\bibitem{semimetal_effects11}
Y. Chen, S. Wu, and A. Burkov,
{\sl ``Axion response in Weyl semimetals''},
Phys. Rev. B {\bf 88}, 125105 (2013) [arXiv:1306.5344].

\bibitem{semimetal_effects12}
Y. Chen, D. Bergman, and A. Burkov,
{\sl ``Weyl fermions and the anomalous Hall effect in metallic ferromagnets''},
Phys. Rev. B {\bf 88}, 125110 (2013) [arXiv:1305.0183];
David Vanderbilt, Ivo Souza, and F. D. M. Haldane
Phys. Rev. B {\bf 89}, 117101 (2014) [arXiv:1312.4200].

\bibitem{semimetal_effects13}
S. T. Ramamurthy and T. L. Hughes,
{\sl ``Patterns of electro-magnetic response in topological semi-metals''},
arXiv:1405.7377.









\bibitem{Zyuzin:2012tv}
A.~A.~Zyuzin and A.~A.~Burkov,
  {\sl ``Topological response in Weyl semimetals and the chiral anomaly,''}  Phys.\ Rev.\ B {\bf 86} (2012) 115133  [arXiv:1206.1868 [cond-mat.mes-hall]].  


\bibitem{tewary}
Pallab Goswami, Sumanta Tewari, {\sl Axionic field theory of (3+1)-dimensional Weyl semi-metals,}
Phys. Rev. B 88, 245107 (2013), arXiv:1210.6352

\bibitem{16}
Chao-Xing Liu, Peng Ye,  Xiao-Liang Qi,
{\sl Chiral gauge field and axial anomaly in a Weyl semimetal},
Physical Review B, vol. 87, Issue 23, id. 235306


\bibitem{VZ}
G.E. Volovik and M.A. Zubkov, "Emergent Weyl spinors in multi-fermion systems," Nuclear Physics B 881, 514 (2014).

\bibitem{W2}
A.A. Soluyanov, D. Gresch, Zhijun Wang, QuanSheng Wu, M. Troyer, Xi Dai, B.A. Bernevig, "Type-II Weyl Semimetals," Nature 527, 495 - 498 (2015).

\bibitem{VolovikBHW2}
 G.~E.~Volovik,
  ``Black hole and Hawking radiation by type-II Weyl fermions,''
  Pisma Zh.\ Eksp.\ Teor.\ Fiz.\  {\bf 104} (2016) no.9,  660
   [JETP Lett.\  {\bf 104} (2016) no.9,  645]
  doi:10.7868/S0370274X16210104, 10.1134/S0021364016210050
  [arXiv:1610.00521 [cond-mat.other]].

\bibitem{NissinenVolovik2017a}
J. Nissinen and G.E. Volovik,
Type-III and IV interacting Weyl points,
Pisma ZhETF {\bf 105}, 442--443 (2017)
JETP Lett.  {\bf 105},  447--452 (2017),
arXiv:1702.04624.

\bibitem{Padmanabhan:2003gd}
  T.~Padmanabhan,
  ``Gravity and the thermodynamics of horizons,''
  Phys.\ Rept.\  {\bf 406} (2005) 49
  doi:10.1016/j.physrep.2004.10.003
  [gr-qc/0311036].



\bibitem{Novikov:1980ni}
  I.~D.~Novikov and A.~A.~Starobinsky,
  ``Quantum Electrodynamic Effects Inside A Charged Black Hole And The Problem Of Cauchy Horizons,''
  Sov.\ Phys.\ JETP {\bf 51} (1980) 1
   [Zh.\ Eksp.\ Teor.\ Fiz.\  {\bf 78} (1980) 3].

\bibitem{Khriplovich:1999gm}
  I.~B.~Khriplovich,
  ``Particle creation by charged black holes,''
  Phys.\ Rept.\  {\bf 320} (1999) 37.
  doi:10.1016/S0370-1573(99)00078-2

\bibitem{Hamilton:2004aw}
  A.~J.~S.~Hamilton and S.~E.~Pollack,
  ``Inside charged black holes. II. Baryons plus dark matter,''
  Phys.\ Rev.\ D {\bf 71} (2005) 084032
  doi:10.1103/PhysRevD.71.084032
  [gr-qc/0411062].

\bibitem{Suuskind}
Leonard Susskind,
"The paradox of quantum black holes",
Nature Physics 2, 665 - 677 (2006)
doi:10.1038/nphys429

\bibitem{Susskind_book}
L. Susskind and J. Lindesay, “An introduction to black holes, information and the string theory revolution: The holographic universe,” Hackensack, USA: World Scientific (2005)
183 p.

\bibitem{Mann}
R. B. Mann, Black Holes: Thermodynamics, Information,
and Firewalls (Springer, New York, 2015)

\bibitem{information}
D. N. Page, “Black hole information,” hep-th/9305040.

\bibitem{BHI}
D. Harlow, “Jerusalem Lectures on Black Holes and Quantum Information,”
Rev. Mod. Phys. 88, 15002 (2016) [Rev. Mod. Phys. 88, 15002 (2016)]
doi:10.1103/RevModPhys.88.015002 [arXiv:1409.1231 [hep-th]].

\bibitem{experiment_BH}
Abbott, B. P. et al. [LIGO Scientific and Virgo Collaborations] GW150914: Implications for the stochastic gravitational wave
background from binary black holes. Phys. Rev. Lett. 116, 131102 (2016).

Johnson, M. D.; Fish, V. L.; Doeleman, S. S.; Marrone, D. P.; Plambeck, R. L.; Wardle, J. F. C.; Akiyama, K.; Asada, K.; Beaudoin, C. (2015-12-04). "Resolved magnetic-field structure and variability near the event horizon of Sagittarius A*". Science. 350 (6265): 1242–1245. Bibcode:2015Sci...350.1242J. ISSN 0036-8075. PMID 26785487. arXiv:1512.01220
doi:10.1126/science.aac7087.

Overbye, Dennis (11 February 2016). "Physicists Detect Gravitational Waves, Proving Einstein Right". New York Times. Retrieved 11 February 2016.


\bibitem{Hamilton:2008zz}
  A.~J.~S.~Hamilton and P.~P.~Avelino,
  ``The Physics of the relativistic counter-streaming instability that drives mass inflation inside black holes,''
  Phys.\ Rept.\  {\bf 495} (2010) 1
  doi:10.1016/j.physrep.2010.06.002
  [arXiv:0811.1926 [gr-qc]].

\bibitem{Lifshitz}
  S.~Liberati, L.~Maccione and T.~P.~Sotiriou,
  ``Scale hierarchy in Horava-Lifshitz gravity: a strong constraint from synchrotron radiation in the Crab nebula,''
  Phys.\ Rev.\ Lett.\  {\bf 109} (2012) 151602
  doi:10.1103/PhysRevLett.109.151602
  [arXiv:1207.0670 [gr-qc]].

\bibitem{Liberati}
  S.~Liberati,
  ``Tests of Lorentz invariance: a 2013 update,''
  Class.\ Quant.\ Grav.\  {\bf 30} (2013) 133001
  doi:10.1088/0264-9381/30/13/133001
  [arXiv:1304.5795 [gr-qc]].

\bibitem{Klinkhamer:2016jrt}
  F.~R.~Klinkhamer and G.~E.~Volovik,
  ``Propagating q-field and q-ball solution,''
  Mod.\ Phys.\ Lett.\ A {\bf 32} (2017) no.18,  1750103
  doi:10.1142/S0217732317501036
  [arXiv:1609.03533 [hep-th]].
\bibitem{Z2016AOP}
  M.~A.~Zubkov,
  ``Wigner transformation, momentum space topology, and anomalous transport,''
  Annals Phys.\  {\bf 373} (2016) 298
  doi:10.1016/j.aop.2016.07.011
  [arXiv:1603.03665 [cond-mat.mes-hall]].


\end{thebibliography}
\end{document}